\documentclass[structabstract]{aa}
\usepackage{graphicx}
\usepackage{natbib}
\usepackage{url}
\usepackage{txfonts}

\begin{document}

   \title{Catalogue of variable stars in open cluster fields}

   \author{Miloslav Zejda
          \inst{1}
          \and
          Ernst Paunzen\inst{1,2,3}
          \and
          Bernhard Baumann\inst{3}
          \and
          Zden\v ek Mikul\'a\v sek\inst{1,4}
          \and
          Ji\v r\'i Li\v ska\inst{1}}

   \institute{Department of Theoretical Physics and Astrophysics, Masaryk University,
              Kotl\'a\v rsk\'a 2, CZ-611~37 Brno, Czech Republic\\
              \email{zejda@physics.muni.cz}
         \and
            Rozhen National Astronomical Observatory, Institute of Astronomy of the Bulgarian
            Academy of Sciences, P.O. Box 136, BG-4700 Smolyan, Bulgaria
         \and
             Institute for Astronomy (IfA), University of Vienna,
              T\"urkenschanzstrasse 17, A-1180 Vienna
         \and
            Observatory and Planetarium of Johann Palisa, V\v SB --
            Technical University, Ostrava, Czech Republic}

   \date{Received 20 March, 2012; accepted 4 October 2012}

  \abstract
   {We present the first catalogue of known variable stars in open cluster regions and with up to two
   times the given cluster radius. This gives basic information about the distribution of variable stars in
cluster fields for the complete sky.}
   {Knowledge of the variable star contents in open clusters is a significant advantage in their
   study. Analysing variability of cluster members and fields stars as well allows us to study
   the characteristics of stars and clusters together.
   This catalogue of variable stars in open cluster fields is the first step in supporting such
   studies.}
   {We took all variable and suspected variable stars into account from the most complete
   collection, ``The AAVSO Variable Star Index'', and did a cross-match of these stars with the
   most complete catalogue of galactic open clusters named DAML02.}
   {Our on-line catalogue presently contains 18\,065 variable stars. We present
   the basic statistical distribution according to types of variability.}
   {}


\maketitle

%

\section{Introduction}

The study of an individual star provides only limited, frequently inaccurate and
uncertain information about it and possibly about the interstellar medium between
this object and us. On the other hand, open clusters provide an ideal opportunity to
simultaneously study a group of stars located in a relatively small space, at the
same distance from the Sun, and with the same age and initial chemical composition.
The detection of any variable star in such stellar aggregates and its usage for
further information mining make research of open clusters very effective. For
example, when observing stellar clusters containing variable stars, it is possible
to obtain data of the characteristics of variable stars, as well as of whole open
clusters \citep{joshi}. This improves our knowledge about both variable stars and
open clusters and yields new data for the study of the dynamics, evolution, and
structure of the whole Galaxy. This idea is partly valid even in the case that an
individual star, the aim of a study, is only located in an open cluster field. In
time of CCDs, it is clear that an observer simultaneously obtains photometric data
about all stars in the corresponding frame. Thus, such observations of a star only
located in an open cluster field, without a confirmed membership in it, can help
study the open cluster itself as well. We are certain that in this context, it is
important to have a catalogue of variable stars in open cluster fields.

A similar project for Galactic globular clusters was initiated by \citet{clement} on
the basis of the first corresponding catalogue by \citet{sawyer}. For globular
clusters, the task of compiling a list of variable members is rather straightforward
because there is almost no contamination of fore- and background objects in the
cluster areas.

At present, about 2100 open clusters or candidates of clusters are known in the
Galaxy (DAML02 Catalogue, Version 3.2, \citealt{dias02a,dias12}), and this is
probably only a low percentage of their total population. Most open clusters are too
far away and therefore too faint to be observable. Furthermore, a substantial
portion of open clusters is hidden behind interstellar material in the Galaxy plane
\citep{froebrich}. From an evolutionary point of view it is obvious that we are
currently only aware of a small portion of all clusters that have ever existed in
our Galaxy. As they drift along their orbits during their lives, some cluster
members escape the hosting cluster thanks to velocity changes in mutual close
encounters (whole clusters or their members), tidal forces in the galactic
gravitational field, and encounters with field stars and interstellar clouds
crossing their way. A typical open cluster has lost most of its member stars after
several hundred Myrs \citep{fuente}. Only some of them reach an age older than a few
billion years. The escaped individual stars continue to orbit the Galaxy on their
own as field stars. This has given rise to the popular view that most or even all
field stars in our Galaxy as well as other galaxies have their origin in clusters.
Thus, the importance of open cluster research is crucial.

Although it seems that the frequency of open cluster studies is increasing, many
clusters remain unstudied, and most important questions still don't have definitive
answers. So far, there are limited surveys like the WIYN Open Cluster Study, which
investigates 14 northern clusters \citep{geller}, or the Southern Open Cluster Study
(SOCS), which includes 24 southern open clusters \citep{kinemuchi}. There are also
several individual studies, such as \cite{march10}, \cite{twar11}, or
\cite{paunzen}. The most complete database of data devoted to open clusters is
WEBDA\footnote{\url{http://webda.physics.muni.cz} \citep{webda}}. It includes
astrometric data in the form of coordinates, rectangular positions, proper motions,
photometric measurements, and spectroscopic data, such as spectral classification,
radial velocities, and rotational velocities. In addition, it contains miscellaneous
types of other data such as membership probabilities, orbital elements of
spectroscopic binaries, and periods of variability for different kinds of variable
stars, including a whole set of bibliographic references. At present, WEBDA includes
information about 1100 open clusters in our Galaxy and in the Small Magellanic
Cloud.

Currently, the presentation of the variable star content within WEBDA is somewhat
inhomogeneous. The corresponding data will only be included in the database if
authors of the relevant papers explicitly state that an object is in an open cluster
field.

The relevant data can be found within WEBDA in the following types:
\begin{itemize}
\item FRQ: collection of multiple frequencies for photometrically variable stars
such as $\delta$ Scuti, $\beta$ Cephei, $\gamma$ Doradus, and slowly pulsating B
stars;
\item IDS: the cross-identifications with double and multiple star components from the
Washington Index of double stars \citep{hartkopf};
\item ORB: orbital elements for spectroscopic binaries;
\item PRD: collection of periods for photometrically variable stars such as
eclipsing binaries, contact binaries, variable stars by rotational modulation,
$\delta$ Scuti type pulsators;
\item SB: numbers of known spectroscopic binaries in each cluster.
\end{itemize}
The individual variables are separately listed on each cluster page in the file type
``Notes''. The latest edition of the General Catalog of Variable Stars (hereafter
GCVS) is also incorporated in the data base and can be fully queried.

However, no complete study exists that shows the distribution of different kinds of
variable stars in open clusters. There are only catalogues of eclipsing or
spectroscopic binaries in open clusters in \cite{popova} and \cite{clausen1}. In
addition there are studies of variable stars in individual open clusters
\citep[e.g.][]{xin}. We compiled the first comprehensive list of variable stars in
open clusters and their close vicinity.


\section{Description of the catalogue}

The latest version of the GCVS contains 45\,835 records \citep{gcvs}. The New
Catalogue of Suspected Variable Stars and its Supplement (\citealt{nsv,nsv2})
contain 14\,811 and 11\,206 stars, respectively. However, the number of known
variable stars is several times higher. At present, the most complete database of
variable stars is managed by the American Association of Variable Star Observers,
AAVSO, as the International Variable Star Index (VSX). It is available on-line at
\url{http://www.aavso.org/vsx} or at the Strasbourg astronomical Data Centre
\citep{vsx}. We used the version available at the CDS on April 29, 2012, containing
209\,285 records (including 575 duplicities). Using the list of stars included in
the VSX and the version 3.2. Catalogue of Open Clusters DAML02
\citep{dias02a,dias12}, we did a match between the two sources of data. We divided
the open clusters into two categories according to their sizes, where the limiting
diameter was 60$\arcmin$. We restricted our sample to clusters with diameters of
less than five degrees, with the exception of the Hyades (Melotte 25). The discarded
clusters are Collinder 173, Collinder 285, Collinder 302, Platais 2, Platais 8,
Platais 9, and Platais 10. For both samples of open clusters, we generated a list of
all suspected variables and variable stars located within the fields of open
clusters. We checked the cluster fields and vicinities up to two times the given
cluster radius. In the first group of 461 open clusters smaller than 60$\arcmin$, we
found 8\,938 variable stars. In the second group of 74 open clusters, we located
9\,127 variable objects. An important concern is proper motion data. We matched all
variable stars of our two samples against the PPMXL catalogue of positions and
proper motions within the ICRS \citep{roeser}. We determined new mean cluster
parameters (see section \ref{apm}). In addition, we used the values from the DAML02
catalogue.

The database of variable stars in both groups of open clusters (Table 1) is
available on-line at the CDS and has following structure:
\begin{itemize}
\item [$\bullet$] VSNo -- Number of star, the prefixes ``S" or ``L" respectively mean the
star belongs to an open cluster smaller or larger than 60$\arcmin$.
\item [$\bullet$] Name -- Name of variable star
\item [$\bullet$] PPMXL -- Identifier of star in PPMXL catalogue
\item [$\bullet$] RAVS -- Right ascension J2000.0 (epoch 2000.0) of variable star in degrees
\item [$\bullet$] DEVS -- Declination J2000.0 (epoch 2000.0) of variable star in degrees
\item [$\bullet$] VSpmRA -- Proper motion of variable star in RA*cos(DE) [mas/yr]
\item [$\bullet$] VSpmDE -- Proper motion of variable star in Declination
[mas/yr]
\item [$\bullet$] VSpmRA\_e -- Mean error of variable star proper motion in VSpmRA*cos(DE) [mas/yr]
\item [$\bullet$] VSpmDE\_e -- Mean error of variable star proper motion in DE [mas/yr]
\item [$\bullet$] Nomeas --  Number of measurements used to derive mean proper motion
\item [$\bullet$] Flag1 -- Taken from \citet{roeser}, where one can find more details. It is a
bitwise or number ($\Sigma 2^i$), where each bit number (i) has the meaning:
\begin{description}
\item $\#$0 (1) = If set, one of the coordinates has an excessively large
               chi square;
\item $\#$1 (2) = Row is from PPMXL \citet{roeser08}. These objects are mostly
               Tycho stars that were masked out of USNO-B.
\item $\#$2 (4) = Row is from PPMXL \citet{roeser08} and replaces a single row
               from USNO-B. This is done when the astrometry from PPMXL
               was better (in terms of error estimates) than the astrometry
               of the corresponding PPMXL object.
\item $\#$3 (8) = Row replaces multiple USNO-B1.0 objects. When PPMX contains
               an object that has more than one counterpart in PPMXL, all
               such counterparts are discarded on the assumption that they
               should have been matched in USNO-B1.0 or result from
               erroneous matches. For these rows, bit$\#$1 is always 1.
\end{description}
\item [$\bullet$] Type -- type of variability, see \url{http://www.aavso.org/vsx/index.php?view=about.vartypes} or \url{http://www.aavso.org/vsx/help/
VariableStarTypeDesignationsInVSX.pdf} for details
\item [$\bullet$] Max\_l -- Limit
flag on magnitude in maximum of brightness
\item [$\bullet$] Max -- Magnitude at maximum, or amplitude of light changes
\item [$\bullet$] Max\_u -- Uncertainty flag on maximum of brightness
\item [$\bullet$] Max\_p -- Passband on maximal magnitude
\item [$\bullet$] Min\_a -- Flag ``(" indicates an amplitude given in next column
\item [$\bullet$] Min\_l -- Limit flag on magnitude in minimum of brightness
\item [$\bullet$] Min -- Magnitude at minimum, or amplitude of light changes
\item [$\bullet$] Min\_u -- Uncertainty flag on minimum of brightness
\item [$\bullet$] Min\_p -- Passband on minimal magnitude
\item [$\bullet$] Epoch -- Epoch of maximum or minimum (HJD)
\item [$\bullet$] Epoch\_u -- Uncertainty flag (:) on epoch
\item [$\bullet$] Period\_l -- Limit flag on period. If the period is lower or higher than limit, then value is
preceded by ``$>$" or ``$<$". ``(" indicates that the period is the mean cycle time
of a U Gem or
     recurrent nova
\item [$\bullet$] Period -- Period of the variable in days
\item [$\bullet$] Period\_u -- Uncertainty or note flag on Period (``:" uncertainty flag, ``)" value of the mean cycle
for U Gem or recurrent nova, ``*N" the real period is likely a multiple of the
quoted period, ``/N" the period quoted is likely a multiple of the real period.
\item [$\bullet$] DVSm -- Distance to the centre of open cluster in arcmin
\item [$\bullet$] DVSp -- Distance to the centre of open cluster in percent of open cluster
apparent radius Diam/2
\item [$\bullet$] OCL -- Name of open cluster (OC)
\item [$\bullet$] RAOC -- Right ascension (J2000.0) of OC centre
\item [$\bullet$] DEOC -- Declination (J2000.0) of OC centre
\item [$\bullet$] Flag2 -- Note to open cluster:
\begin{itemize}
\item \textit{e} embedded open cluster (or cluster associated with nebulosity)
\item \textit{g} possible globular cluster
\item \textit{m} possible moving group
\item \textit{n} ``non-existent NGC" (RNGC, \citealp{sulentic}). Some of Bica's POCRs are also ``non-existent NGC" objects.
\item \textit{o} possible OB association (or detached part of)
\item \textit{p} POCR (Possible Open Cluster Remnant) - \citealp{bica}
\item \textit{v} clusters with variable extinction \citep{ahumada}
\item \textit{r} recovered: ``non-existent NGC" that are very visible in the DSS image inspection
\item \textit{d} dubious: objects considered doubtful by the DSS image inspection
\item \textit{nf} not found: objects not found in the DSS image inspection (wrong coordinates?)
\item \textit{cr} cluster remnant \citep{pavani}
\item \textit{OC?} possible cluster (Kronberger M., private communication)
\item \textit{OC} likely cluster (Kronberger M., private communication)
\item \textit{IR} discovered in infra-red but are visible in the DSS images inspection.
\end{itemize}
\item [$\bullet$] Diam -- Apparent diameter of open cluster in arcmin
\item [$\bullet$] \textit{d} -- Distance of open cluster [pc]
\item [$\bullet$] \textit{E(B-V)} -- reddening, colour excess in \textit{(B-V)} [mag]
\item [$\bullet$] $\log t$ -- logarithm of the age of open cluster in years
\item [$\bullet$] Nc -- estimated number of members in the open cluster
\item [$\bullet$] OCpmRA -- mean cluster proper motion in RA [mas/yr]
\item [$\bullet$] OCpmRA\_e -- error of mean cluster proper motion in $\mu_{\alpha}$.$\cos \delta$, ICRS [mas/yr]
\item [$\bullet$] OCpmDE -- proper motion in $\mu_{\delta}$, ICRS [mas/yr]
\item [$\bullet$] OCpmDE\_e -- error of proper motion in DE [mas/yr]
\item [$\bullet$] Npm --  number of stars used to derive mean cluster proper motion
\item [$\bullet$] Ref1 -- reference to proper motion -- \textit{Z} -- this paper, Table 2;
          \textit{D} - see \cite{dias02a} for details
\item [$\bullet$] RV -- mean cluster radial velocity [km.s$^{-1}$]
\item [$\bullet$] eRV -- error of radial velocity [km.s$^{-1}$]
\item [$\bullet$] Nrv - number of stars used to derive mean cluster radial velocity
\item [$\bullet$] Ref2 -- reference to radial velocity -- see \cite{dias02a} for details
\item [$\bullet$] ME -- mean cluster metallicity
\item [$\bullet$] eME -- error of cluster metallicity
\item [$\bullet$] Nme -- number of stars used to determine mean cluster metallicity
\item [$\bullet$] TrTyp -- Trumpler classification of open cluster.
\end{itemize}

The number of variable stars is increasing rapidly. The catalogue given in Table 1
is only the starting version of such a catalogue, which should be updated. We are
going to publish updates irregularly when the number of newly added variable stars
in VSX reaches 15\,000.

\subsection{Mean proper motions of open clusters} \label{apm}

A homogeneous data set of mean cluster proper motions has not been available until now.
We therefore used the following sources (sorted alphabetically) to compile a new
catalogue:
\begin{itemize}
    \item \citet{baumgardt}: based on the Hipparcos catalogue \citep{hipper}
    \item \citet{beshenov}: based on the Tycho-2 catalogue \citep{hog}
    \item \citet{dias01,dias02b}: based on the Tycho-2 catalogue
    \item \citet{dias06}: based on the UCAC2 catalog \citep{zacharias}
    \item \citet{frinch}: based on the Tycho-2 catalogue
    \item \citet{kharchenko05}: based on the ASCC2.5 catalogue \citep{kharchenko01}
    \item \citet{krone}: based on the Bordeaux PM2000 proper motion catalogue
    \citep{ducourant}
    \item \citet{robichon}: based on the Hipparcos catalogue
    \item \citet{leeuwen}: based on the new Hipparcos catalogue
\end{itemize}
If more than one measurement is available, a weighted mean was calculated, using the
reciprocal errors listed in the individual source as weights. Outliers were
identified when more than two values were available. Only seven outliers were found
in the complete sample: Harvard~76, IC~4665, Melotte~111, NGC~7092, Pismis~28 and
Ruprecht~9 \citep{dias06}, NGC 6250 \citep{frinch}, as well as Melotte~20
\citep{dias01}. Those data points were not included in the final compilation.

In total, a catalogue of proper motions for 879 open clusters was compiled, from
which 436 have more than one available measurement. Table 2 lists the catalogue.

\setcounter{table}{1}
\begin{table}
\begin{center}
\caption[]{New compilation of proper motions for 721 open clusters
($\mu_{\alpha^{\ast}}$\,=\,$\mu_{\alpha}\cdot\cos\delta$). The complete table is
only available on-line.}
\begin{tabular}{@{}lrcrcr@{}}
\hline Cluster & $\mu_{\alpha^{\ast}}$ & $\sigma(\mu_{\alpha^{\ast}})$ & $\mu_{\delta}$ &
$\sigma(\mu_{\delta})$ & $N$ \\
\hline
Alessi~1    &   +6.5    &   0.4 &   $-$7.7  &   0.3 &   1   \\
Alessi~2    &   $-$0.1  &   0.4 &   +0.3    &   0.5 &   1   \\
Alessi~3    &   $-$11.6 &   0.4 &   +12.6   &   0.5 &   1   \\
Alessi~5    &   $-$9.9  &   0.3 &   +3.6    &   0.4 &   2   \\
Alessi~6    &   $-$6.0  &   0.4 &   $-$5.7  &   0.4 &   2   \\
Alessi~8    &   $-$5.3  &   0.3 &   $-$5.4  &   0.3 &   2   \\
Alessi~9    &   +11.6   &   0.3 &   $-$9.7  &   0.3 &   1   \\
Alessi~10   &   +1.3    &   0.4 &   $-$8.6  &   0.3 &   2   \\
Alessi~12   &   +1.5    &   0.1 &   $-$4.0  &   0.2 &   2   \\
Alessi~13   &   +37.0   &   0.7 &   $-$3.2  &   0.4 &   1   \\
Alessi~19   &   $-$0.5  &   0.2 &   $-$7.0  &   0.4 &   1   \\
Alessi~20   &   +7.5    &   0.4 &   $-$2.6  &   0.5 &   1   \\
Alessi~21   &   $-$3.7  &   0.4 &   +3.0    &   0.4 &   1   \\
\hline
\end{tabular}
\label{apm_list}
\end{center}
\end{table}


\section{Results, statistics}

\subsection{Membership}

A detailed membership analysis of the catalogue stars to the corresponding cluster
is beyond the scope of this paper. However, as a first heuristic approximation, we
present the dependence of areal density of variable stars on the distance to the
published cluster centres. The distances are expressed in units of the given
accepted cluster radii. We want to emphasise that, although several improvements
have been made over the last decades, most of the centre coordinates and angular
radii still arise from the visual inspection of photographic plates.

\begin{figure}
\begin{center}
\includegraphics[scale=0.33]{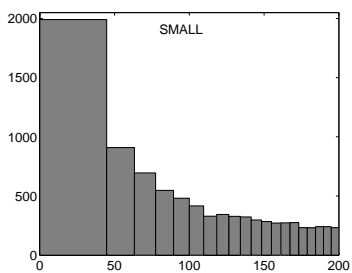} \caption{Distribution of variable
stars according to the distance to the cluster centre in units of the corresponding
radius for open clusters of diameters smaller than 60\arcmin. }\label{fig:tot}
\end{center}
\end{figure}

Figure \ref{fig:tot} shows the ``distance distributions'' only for studied open
clusters smaller than 60\arcmin. We constructed a system of concentric circles
around the centre of each open cluster with the same area. Thus we can really study
the areal density of variable stars in increasing distant from the open cluster
centre. In this figure and fig. \ref{fig:diag2} columns with decreasing widths
correspond to a decreasing thickness of those concentric circles. We learned that
variable stars found in the area of open clusters larger than 60\arcmin (mostly
nearby clusters) are strongly dominated by variables behind them. However, a
detailed kinematic study of the objects might overcome this problem because these
clusters should have significant proper motions, while background objects have none.
The large all-sky-catalogues, like the PPMXL \citep{roeser}, still lack the accuracy
needed to perform such analysis.

For the smaller clusters we can assume that stars closer than 50 \% to the centre of
the cluster are most probably members of the cluster according to the typical shape
of cluster density profiles \citep{seleznev}. In general, the radial density for the
members of open clusters can be well fitted by a King profile. \citet{sanchez}
present the internal spatial structure of 16 open clusters on the basis of
positional, as well as kinematic data. Figure 2 (left panel) shows the distribution
of members for the distant open cluster NGC~2194, which has a diameter of 9\arcmin.
Concentration of the members to the centre is clearly visible, supporting our
consideration. About 60\% of our corresponding objects fall within this range. Going
further out from the centre, the density of variable stars drops to the density of
the field variable stars.

\subsection{Type of variability}

\begin{figure}
\begin{center}
\includegraphics[scale=0.33]{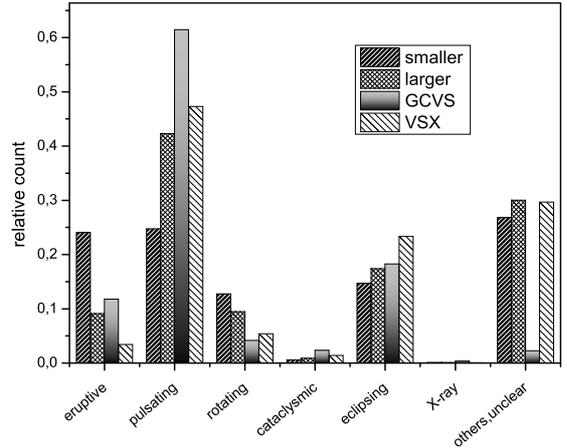}
\caption{Relative numbers of basic star variability types in open clusters smaller and larger than
60$\arcmin$, compared to the GCVS and VSX. The values are normalised to the number of variable
stars in the given source of data.}\label{fig:count}
\end{center}
\end{figure}

There are about almost one hundred different classes and subclasses of variability given
in the GCVS. However, new types and new divisions of variable stars still appear. The new
structure was proposed at the 26th General Assembly of the IAU \citep{samus06}, but it
has not been accepted so far. The Variable Star Index of the AAVSO very often uses ``new
definitions'' of variability types taken from original papers, especially when describing
results of a survey. In this paper we used types of variability from AAVSO
VSX\footnote{see details in help at
\url{http://www.aavso.org/vsx/help/VariableStarTypeDesignationsInVSX.pdf} and GCVS
\citep{gcvs}.} However, we used the original GCVS division in the following categories of
variable stars: eruptive, pulsating, rotating, cataclysmic variables, eclipsing binary
systems, intense variable X-ray sources, and others. The group of new variability types
was dissolved into the previous seven groups according to the main characteristics of the
objects according to variable star designation in VSX. Objects with unknown kinds of
variability are marked as ``--'', ``*'', VAR, or MISC with no information about
variability were counted into the group ``Others''. Unclear classification of variability
shown by a sign ``:'' is counted in the corresponding group and separately as an unclear
member of this group. When the classification contains two or more types with a logical
``or'' (sign ``$\mid$''), the classification is uncertain and all possible types are
indicated. If these types belong to one class of variability we count the star once in
that type of variability. If classified types differ we count them separately as unclear
determination in corresponding groups of variability. When a star shows two types of
variability which is demonstrated by a logical ``and'' (sign ``+''), we count the star in
each of the groups of variable stars. That certainly means that the total sum of stars in
all groups is larger than the number of stars in the catalogue. The obtained distribution
of the main types of variable stars in both catalogues (VSX and GCVS) and for both
smaller and larger open clusters are given in Tables \ref{table:2} and \ref{table:3} and
shown in Figure \ref{fig:count}. The given variable star-type distribution provides us
only a raw image of the real situation. Here, the compared GCVS sample is the electronic
edition of March 2012 at the CDS \citep{gcvs}, which contains 45\,835 variable stars. The
used AAVSO Variable Star Index\footnote{\url{http://www.aavso.org/vsx}} is the public
version available at CDS on April 29, 2012 containing 209\,285 objects \citep{vsx}. The
number of unveiled variable stars is growing, and the classification of variability type
is improving, but unfortunately the processing is not going evenly. The sky coverage is
then not proportioned well.

\begin{table}[h] \caption{Numbers of different type variable stars in open
clusters smaller and larger than 60$\arcmin$. }             
\label{table:2}      
\begin{center}
\begin{tabular}{l|rrr|rrr|r}        
\hline\hline  
Type          & \multicolumn{3}{l}{Smaller}  & \multicolumn{3}{l}{Larger} & Total \\
\hline
eruptive        & 1\,982 & 167 &   0 &  749 &  83 &   0 & 2\,981  \\
pulsating       & 1\,819 & 303 &  90 & 3\,112 & 539 & 209 & 6\,072  \\
rotating        & 1\,018 & 118 &   3 &  793 &  61 &  12 & 2\,005  \\
cataclysmic     &   41 &  11 &   0 &   68 &  11 &   1 &  132  \\
eclipsing       & 1\,162 & 110 &  42 & 1\,404 &  86 & 100 & 2\,904  \\
X-ray sources   &    7 &   1 &   0 &    6 &   0 &   1 &   15 \\
unclear, others & 1\,948 & 388 &  60 & 2\,432 & 174 & 133 & 5\,135 \\
\hline
\end{tabular}
\end{center}
Three columns are given for both cluster types larger and smaller, respectively.
The first column contains number of star with clear classification of variability
type, the second column numbers of stars with uncertain classification denoted by a
sign ``:''. The third column numbers of unclearly classified stars noted by a sign
``$\mid$''. The last column of the table contains sums of all stars of that
variability class including unclear and uncertain classified ones.
\end{table}

\begin{table}[h]
\caption{Numbers of types of variable stars in present variable star catalogues in the
same form as explained in Table \ref{table:2}. }             
\label{table:3}      
\centering                          
\begin{tabular}{l|rrr|rrr}        
\hline\hline                 
Type          & \multicolumn{3}{l}{GCVS} & \multicolumn{3}{l}{VSX}  \\
\hline
eruptive        &  4\,781 &  606& 0 &    6\,005&  1\,098 &   12  \\
pulsating       & 23\,048 & 5\,120& 0 &   78\,970& 13\,254 & 6\,784  \\
rotating        &  1\,434 &  468& 0 &   10\,521&   659 &  148  \\
cataclysmic     &   846 &  221& 0 &    2\,406&   486 &   23  \\
eclipsing       &  7\,616 &  748& 0 &   44\,083&  1\,621 & 3\,123  \\
X-ray sources   &   157 &    8& 0 &     103&     8 &    2 \\
unclear, others &   757 &  280& 0 &   54\,335&  3\,591 & 4\,142 \\
\hline                                   
suma of stars & \multicolumn{3}{l}{45\,835} & \multicolumn{3}{l}{209\,285} \\
\end{tabular}
\end{table}

\begin{figure}
\begin{center}
\includegraphics[scale=0.28]{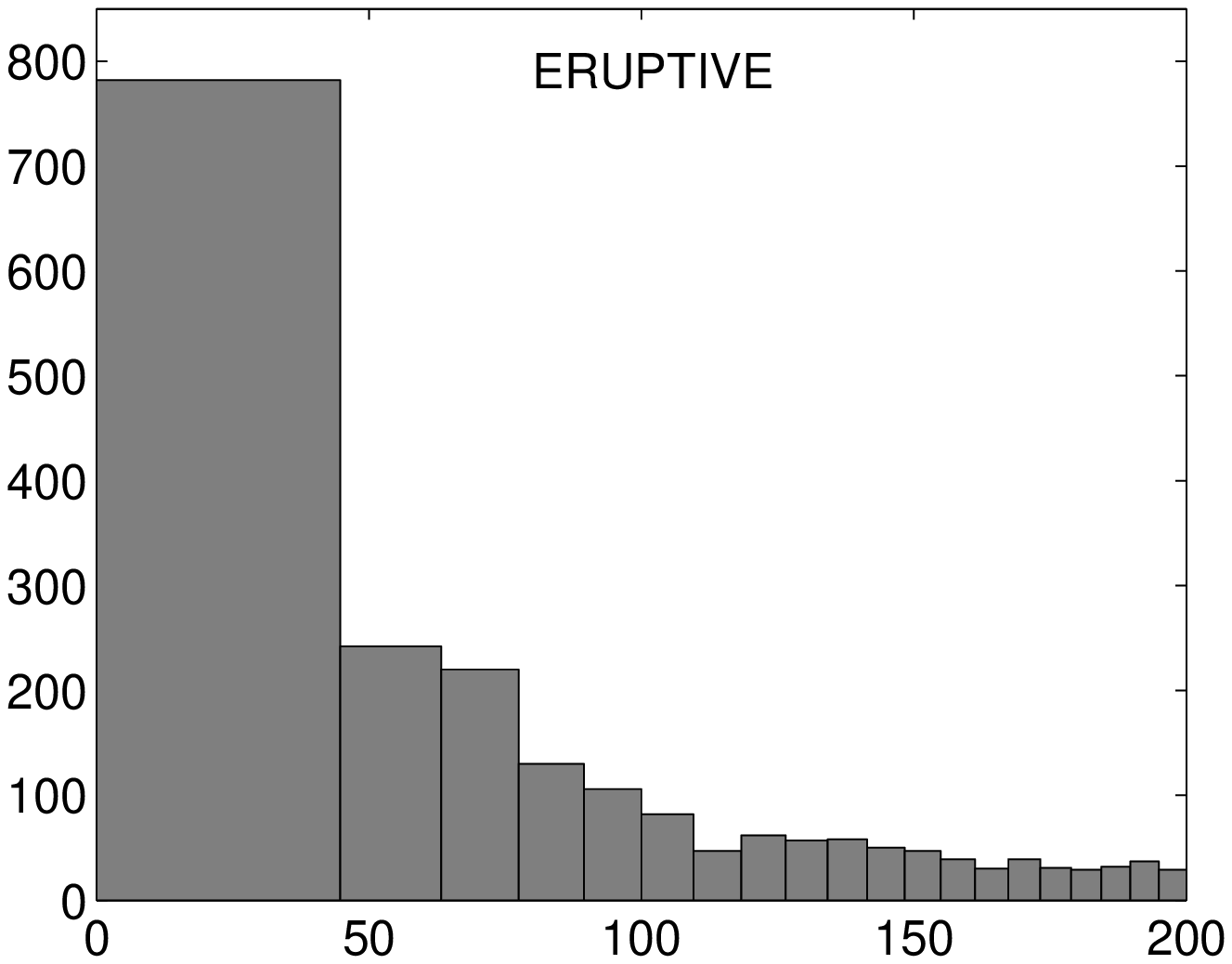}
\includegraphics[scale=0.28]{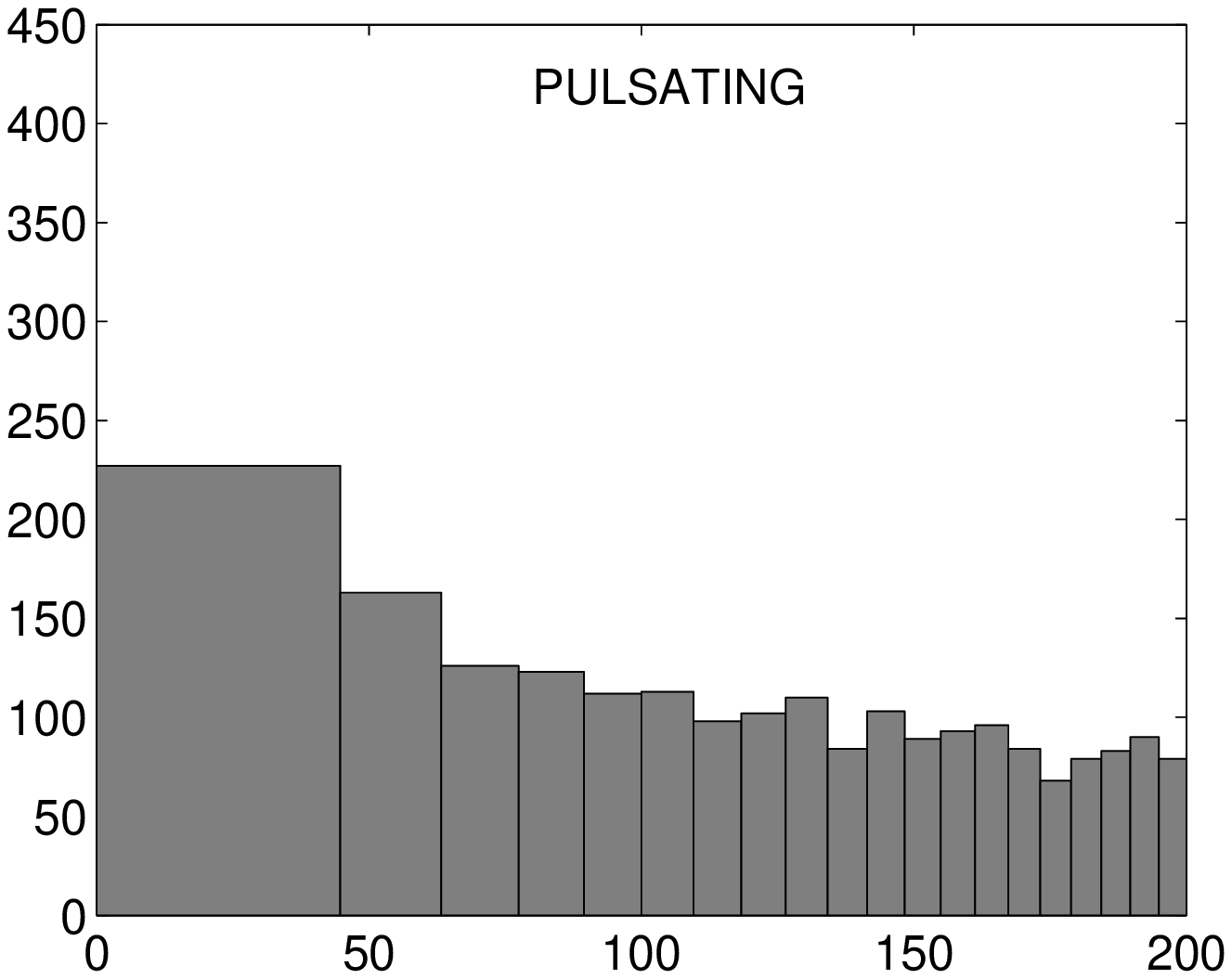}\\
\includegraphics[scale=0.28]{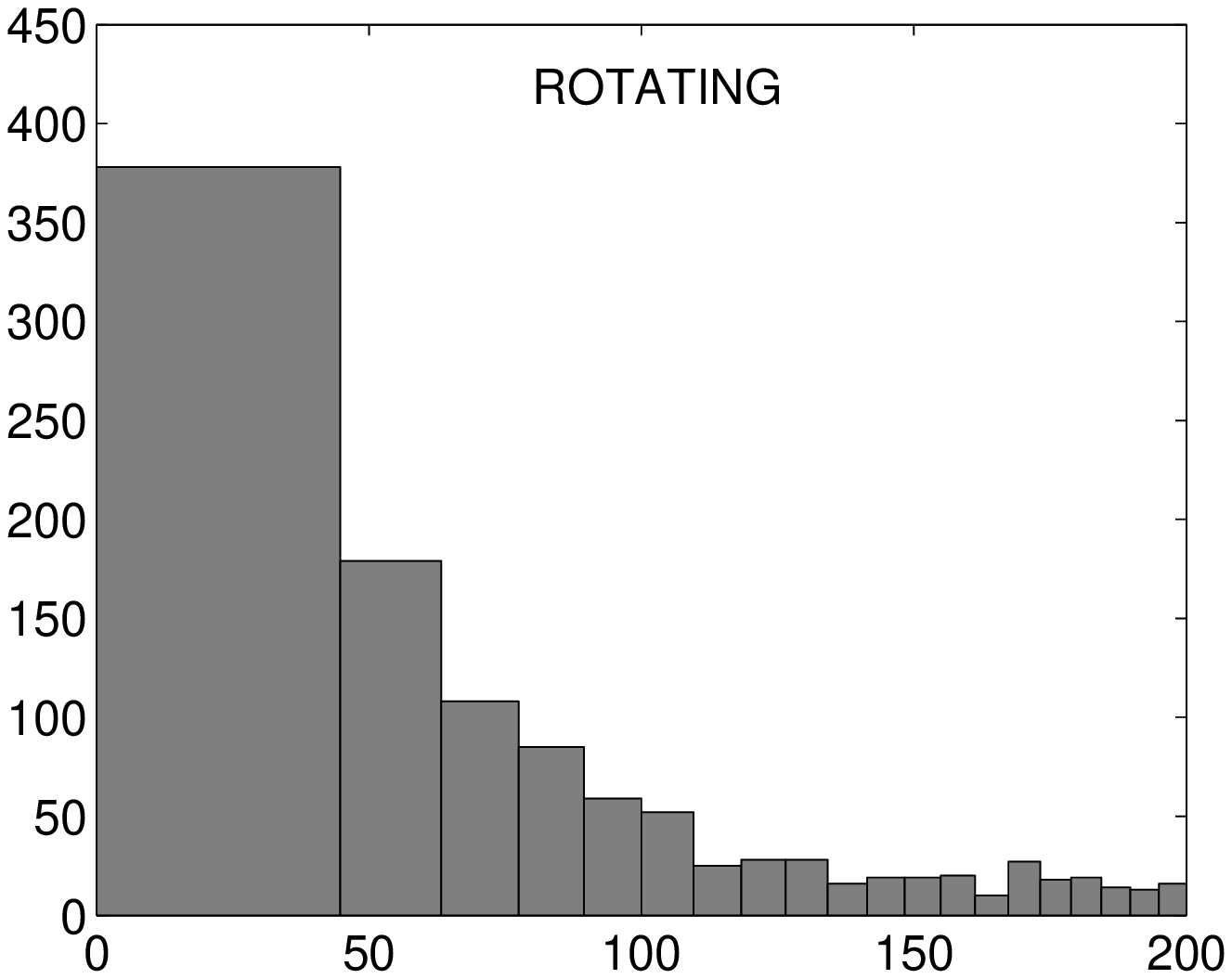}
\includegraphics[scale=0.28]{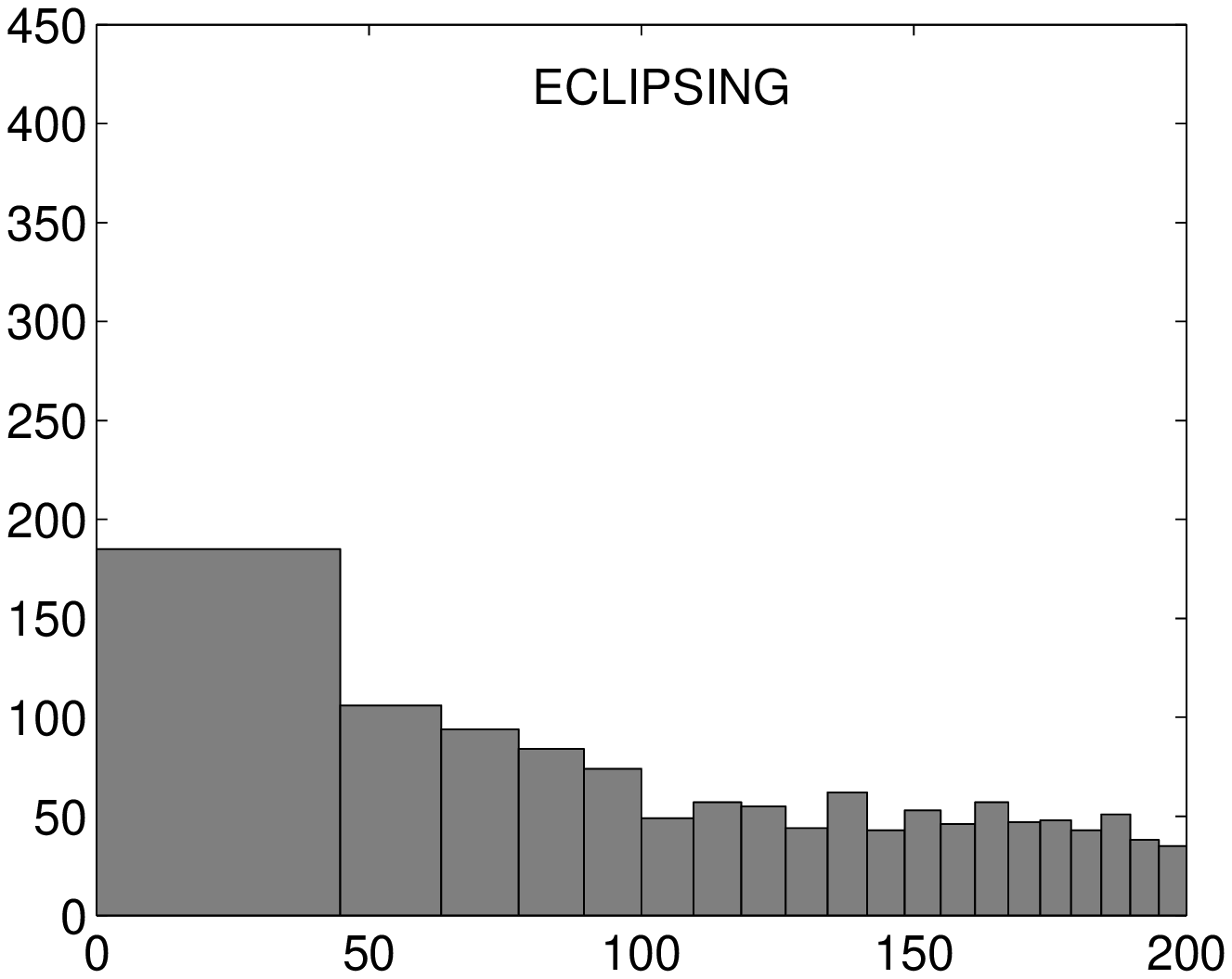}\\
\caption{Distribution of the four most numerous groups of variable stars according to their
distance from the centre of open clusters in cluster radii in open clusters smaller than
60$\arcmin$ in diameter.}\label{fig:diag2}
\end{center}
\end{figure}

After inspecting Fig.\,\ref{fig:diag2} we conclude that the four most numerous
groups of variable stars in open clusters all display an apparent concentration
around the centre of small clusters. However, the affinity is different. Ignoring
the most numerous group of yet unconfirmed or unspecified variables, the most common
types of variables found in the area within two radii from the centres of open
clusters are eruptive and pulsating variables (see\,Table \ref{table:3}). They have
quite a disjuncted distribution function: while the prevailing majority of eruptive
variables is concentrated around the centres of open clusters, the pulsating stars
are distributed much more evenly, which can be interpreted as a significant portion
of these pulsating stars not belonging to the open clusters themselves. This
interpretation is supported by the distribution of pulsating variables for larger
open clusters, which is almost the same (182 variables) for all areas. We also
inspected the distribution of the other types of star variability, rotating stars
and eclipsing binaries (see Fig. \ref{fig:diag2}). The distribution of rotating
variables indicates that practically all these observed variables belong to
clusters, while the distribution of eclipsing binaries resembles that of the
distribution of pulsating stars with some relation to their ``host'' clusters.

However, some new photometric studies of open clusters \citep[e.g.][]{sanquist,
janik, lata} show that many variable stars in open clusters still need to be
unveiled, so these statistical results could be significantly distorted due to the
selection of only currently known variables.

\section{Conclusions}

We have prepared a new catalogue of variable and suspected variable stars in open
clusters and their close vicinities. This version of the catalogue is only the start
of the project. The compiled lists of objects include stars that are located in the
area of open clusters up to twice their published radii. Owing to the lack of
accurate and homogeneous data, we were not able to extract detailed membership
probabilities for all stars. However, we took a first step by including the proper
motion of targets from the PPMXL catalogue, and we prepared a new catalogue of mean
cluster proper motions.

The most numerous group of variable stars present in open clusters is the still
unconfirmed or unspecified objects followed by the eruptive variables. The pulsating
variables that dominate the cluster variable star population in open clusters larger
than 60$\arcmin$ are mostly field stars.

We are convinced that our catalogue with noted variability will significantly contribute towards
further research on related topics. For example, we are able to determine the distance of an open
cluster in several ways using variable stars (eclipsing binaries and/or pulsating stars). Thus,
analysing one type of cluster target(s) will have a major impact on similar objects in the galactic
field.

\begin{acknowledgements}
Part of this work was supported by the Czech grants GAP209/12/0217, 7AMB12AT003, WTZ CZ
10/2012, MUNI/A/0968/2009 and Austrian Research Fund via the project FWF P22691-N16. We
appreciate the using of VSX Index of AAVSO. This research has made use of NASA's
Astrophysics Data System and the WEBDA database, operated at the Department of
Theoretical Physics and Astrophysics of the Masaryk University. We thank S. de Villiers
for valuable notes to the manuscript.
\end{acknowledgements}

\end{document}